\documentclass[sigconf]{acmart}

\AtBeginDocument{%
  \providecommand\BibTeX{{%
    \normalfont B\kern-0.5em{\scshape i\kern-0.25em b}\kern-0.8em\TeX}}}

\setcopyright{acmcopyright}
\copyrightyear{2018}
\acmYear{2018}
\acmDOI{10.1145/1122445.1122456}

\acmConference[WWW'2022]{WWW '22: The Web Conference }{April 25-29, 2022}{Online}
\acmBooktitle{WWW'2022: The Web Conference, April 25-29, 2022, Online}
\acmPrice{15.00}
\acmISBN{978-1-4503-XXXX-X/18/06}

\copyrightyear{2022} 
\acmYear{2022} 
\setcopyright{acmcopyright}\acmConference[WWW '22]{Proceedings of the ACM Web Conference 2022}{April 25--29, 2022}{Virtual Event, Lyon, France}
\acmBooktitle{Proceedings of the ACM Web Conference 2022 (WWW '22), April 25--29, 2022, Virtual Event, Lyon, France}
\acmPrice{15.00}
\acmDOI{10.1145/3485447.3511970}
\acmISBN{978-1-4503-9096-5/22/04}
\begin{document}

\title{
Deep Interest Highlight Network for Click-Through Rate Prediction in Trigger-Induced Recommendation
}









\author{Qijie Shen$^{1}$, Hong Wen$^{1}$, Wanjie Tao$^1$, Jing Zhang$^{2}$, Fuyu Lv$^1$, Zulong Chen$^1$, Zhao Li$^3$}
\authornote{Q. Shen and H. Wen share the co-first authorship.}

\affiliation{\institution{$^1$Alibaba Group, $^2$The University of Sydney,$^3$Zhejiang University}\country{}} 

\email{{qijie.sqj, qinggan.wh, wanjie.twj, zulong.czl, fuyu.lfy}@alibaba-inc.com} \email{jing.zhang1@sydney.edu.au,zhao_li@zju.edu.cn} 

\def\authors{Qijie Shen, Hong Wen, Wanjie Tao, Jing Zhang, Fuyu Lv, Zulong Chen, Zhao Li}

\renewcommand{\shortauthors}{Shen, et al.}

\begin{abstract}

In many classical e-commerce platforms, personalized recommendation has been proven to be of great business value, which can improve user satisfaction and increase the revenue of platforms. In this paper, we present a new recommendation problem, Trigger-Induced Recommendation (TIR), where users' instant interest can be explicitly induced with a trigger item and follow-up related target items are recommended accordingly. TIR has become ubiquitous and popular in e-commerce platforms. In this paper, we figure out that although existing recommendation models are effective in traditional recommendation scenarios by mining users’ interests based on their massive historical behaviors, they are struggling in discovering users’ instant interests in the TIR scenario due to the discrepancy between these scenarios, resulting in inferior performance. To tackle the problem, we propose a novel recommendation method named Deep Interest Highlight Network (DIHN) for Click-Through Rate (CTR) prediction in TIR scenarios. It has three main components including 1) User Intent Network (UIN), which responds to generate a precise probability score to predict user's intent on the trigger item; 2) Fusion Embedding Module (FEM), which adaptively fuses trigger item and target item embeddings based on the prediction from UIN; and (3) Hybrid Interest Extracting Module (HIEM), which can effectively highlight users' instant interest from their behaviors based on the result of FEM. Extensive offline and online evaluations on a real-world e-commerce platform demonstrate the superiority of DIHN over state-of-the-art methods. Our code is available \footnote{https://github.com/EzailShen/WWW-22-DIHN}.


\end{abstract}

\ccsdesc[500]{Information system~Information retrieval}

\keywords{Recommender System, Click-Through Rate Prediction, Trigger-Induced Recommendation, Users' Behavior Modelling}


\maketitle

\section{Introduction}

\begin{figure}
  \centering
  \includegraphics[width=\linewidth]{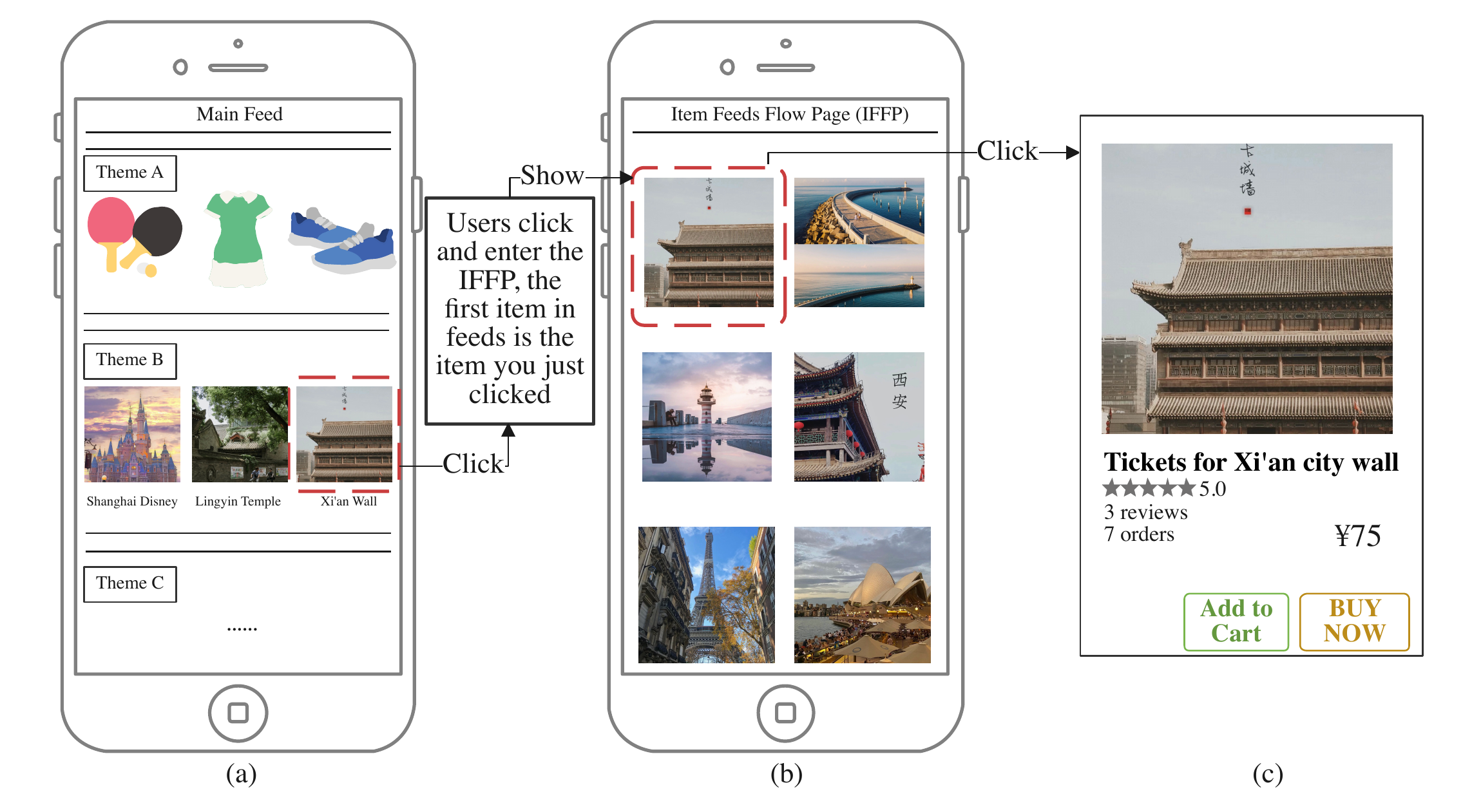}
  \caption{ 
  An illustration of the online TIR scenario in a popular e-commerce platform, where users' instant interest can be explicitly induced by the trigger item (clicked item) and relevant target items are recommended accordingly.   }
  \label{fig:scene_info}
\end{figure}

In traditional recommender system (RS), users can only receive information in a passive manner, lacking of instant feedback mechanisms for interacting with the system. However, users sometimes may intend to actively access more related items with the clicked item just now. To achieve this goal, in this paper, we present a new recommendation problem, Trigger-Induced Recommendation (TIR), where users' instant interest can be explicitly induced with a trigger item ($i.e.$, the last clicked item) and follow-up related items are recommended accordingly. For example in our online scenario (one of the 
largest e-commerce platform in the world), multiple cards with each three recommended items are displayed first as shown in Fig. \ref{fig:scene_info}(a). Then, once an item is clicked, a new page, called Item Feeds Flow Page (IFFP), will be triggered and displayed as shown in Fig. \ref{fig:scene_info}(b), where the clicked item just now, called trigger item, is always located at the first position (highlighted by the red rectangle), and other items are recommended accordingly. Next, if users click one interested item within IFFP, the corresponding detailed page will be displayed as shown in Fig. \ref{fig:scene_info}(c), where users can directly purchase or add it into the Shopping Cart. We refer to this recommendation scenario within IFFP as TIR scenario, which has become ubiquitous and popular in e-commerce platforms. In our online App, TIR has already been the standard recommended scenario, serving an essential way for users to enter the detail page for further purchase. Besides, TIR scenarios have contributed more than 60\% of the item page view (IPV) among all recommendation scenarios from our app, which indicates the significance value of the proposed task and the performance improvements. The similar scenario can also be found in messaging APPs, $e.g.$, recommendation suggestion for relevant passages in WeChat Top Stories~\cite{xie2021real}. In this paper, we focus on the Click-Through Rate (CTR) prediction task for TIR in e-commerce scenarios, which can improve user experience and increase the revenue of the platform.

As we know, CTR is playing a crucial role in recommendation~\cite{zhou2018deep, zhou2019deep,song2020towards,feng2019deep,pi2019practice,shen2021sar}, which aims to predict the probability of users clicking items.
Recently, inspired by the success of deep learning in various research fields, $e.g.$, natural language processing \cite{li2017deep,guo2020gluoncv,zhang2020adversarial} and computer vision \cite{gustafsson2020evaluating,voulodimos2018deep,ioannidou2017deep}, deep learning based methods also have been proposed for the CTR prediction task, such as PNN \cite{qu2016product}, DeepFM \cite{guo2017deepfm}, and DCN \cite{wang2017deep}. These methods, from the perspective of feature interactions, firstly map large-scale sparse features into fixed low dimensional embedding vectors and then feed them into fully connected layers to learn feature representations, neglecting of capturing users' interests from their historical behaviors. Alternatively, increasing novel methods have been proposed to extract users' interests from their historical behaviors. Deep Interest Network (DIN) \cite{zhou2018deep}, the first attention based work in the CTR area, employs attention mechanism to dynamically re-weigh users'
historical behaviors with respect to the target item. Deep Interest Evolution Network (DIEN) \cite{zhou2019deep} is further proposed to model users' interests evolving process in e-commerce system. In Search-based Interest Model (SIM) \cite{pi2020search}, a kind of hard-search mode is proposed to extract users' interests with respect to the target item, which selects and aggregates only behaviors with the same category as the target item into a sub behavior sequence.

Despite effective, we figure out that those methods are not optimal in a trigger-induced recommender due to the discrepancy between TIR and traditional scenarios. Users probably have multiple interests in their historical behaviors, such as electronics, clothing, and snacks, resulting in diverse items recommended accordingly in non-TIR scenarios, $e.g.$, the scenario in Fig. \ref{fig:scene_info}(a). 
However, in TIR scenarios, users' instant interest can be explicitly induced with a trigger item. For example, given the category of the trigger item \emph{Electronic}, it implies that the user is only interested in items related to \emph{Electronic} category at that transient moment. Therefore, if directly employing existing methods, such as DIN \cite{zhou2018deep} or DIEN \cite{zhou2019deep}, the performance will be seriously degraded because they do not explicitly model the instant interest induced by the trigger item. An intuition is that if we can leverage the trigger item to discover users' instant interest, we can recommend more relevant target items that will be probably clicked, $i.e.$, improving the CTR performance. However, it is not trivial due to the following challenges.

\begin{itemize}


\item Challenge 1: Users' instant interest induced from a trigger item are inherently noisy, because there are some accidental clicks on wrong items in users' behaviors. How to evaluate users' real intent for the trigger item remains challenging. 

\item Challenge 2: Users always show multiple interests from their historical behaviors. However, in TIR scenarios, users usually show their instant interest on the trigger item. Therefore, how to extract user's interest from their historical behaviors with respect to the clicked trigger item and the target item simultaneously is unexplored.

\end{itemize}

To address these challenges, we propose a novel model named Deep Interest Highlight Network (DIHN) for CTR prediction in TIR scenarios. Specifically, it consists of a User Intent Network (UIN), a Fusion Embedding Module (FEM), and a Hybrid Interest Extracting Module (HIEM). UIN responds to generate a precise probability score to predict user's intent on the trigger item. FEM adaptively fuses trigger item and target item embeddings based on the prediction from UIN. For example with two extreme cases, one is that if users have no any interest on trigger item but only click it casually, the results of FEM will degrade to the target item embedding. The other is that if users have exactly intense interest on the trigger item, the FEM will degrade to the trigger item embedding. Actually, the result predicted by UIN reflects the intensity of users' instant interest, which is leveraged in FEM to fuse both embeddings in an adaptive manner for better users' interest extraction. 
In addition, HIEM can effectively highlight users' exact interest from their behaviors by leveraging two kinds of modelling paradigms named Soft Sequential Modelling (SSM) and Hard Sequential Modelling (HSM).
SSM adaptively extracts the representation vector of users' interests by taking into consideration the relevance of historical behaviors with respect to the FEM. While HSM, borrowing the idea of hard-search mode from SIM \cite{pi2020search}, firstly selects the behaviors with the same category as the trigger item, and then aggregates theses behaviors as a sub behavior sequence, followed by employing it to capture users' corresponding interests. Finally, all the representation features are concatenated with several raw context features and fed into fully connected layers to generate the final prediction results. 

The main contributions of our paper are as follows:

\begin{itemize}
\item To the best of our knowledge, this is the first work to study the important recommendation problem named trigger-induced recommendation in e-commerce, which poses new challenges beyond existing recommender system.


\item We propose a novel CTR model named DIHN for TIR scenarios, which can learn more expressive user interest representation and achieve more precise CTR prediction based on three carefully devised components. 

\item We conduct extensive experiments on the real-world offline datasets, and the results demonstrate the effectiveness of the proposed DIHN than representative state-of-the-art solutions. It is notable that DIHN has been deployed in our online recommender system and delivers significant improvement, further confirming its value in industrial applications. 
\end{itemize}

The rest of our paper is organized as follows. Section 2 presents a brief review of related works, followed by the details of the proposed model in Section 3. Experiment setups as well as the corresponding results and analysis are presented in Section 4. We finally conclude the paper and discuss the future work in Section 5.

\section{Related work}

In this section, we will review the related work from following three respects briefly: Feature Interaction, User Behavior Modelling and Trigger-Induced Recommendation.

\textbf{Feature Interaction:} Nowadays, academic and industrial communities have paid more and more attention on capturing feature interactions instead of exhausting feature engineering works. DCN \cite{wang2017deep} and Wide\&Deep \cite{cheng2016wide} creatively replace the manual features transformation with neural networks for better memorization and generalization abilities. NCF \cite{he2017neural}, DeepFM \cite{guo2017deepfm} and DMF \cite{xue2017deep} impose a neural network with multiple MLPs to model the feature interactions between users and items. AutoInt \cite{song2019autoint} and CAN \cite{zhou2020can} further propose the  self-attention mechanism for comprehensive feature interactions. Moreover, GNNs \cite{gori2005new} realize feature interaction from the perspective of graphs and achieve great success. In our DIHN, we also explore higher-order feature interactions from all the feature embeddings as well as raw context features.


\textbf{User Behavior Modelling:} Recently, a series of works are proposed to capture users' interests from their rich historical behavior data with different neural network architecture such as Transformer \cite{feng2019deep, sun2019bert4rec}, CNN \cite{tang2018personalized, yuan2019simple}, Capsule \cite{li2019multi}, RNN \cite{hidasi2015session}. For example, DIN \cite{zhou2018deep} emphasizes that users' interests are diverse and an attention mechanism is introduced to capture users’ diverse interests on the different target items. DIEN \cite{zhou2019deep} refines GRU to model evolution of interest and proposes an auxiliary loss to capture latent interest from users' behaviors. Pi et al. \cite{pi2020search} proposes a novel memory based architecture named MIMN to capture users' interests from long sequential behavior data, which is the first industrial solution that is capable of handling long sequential user behavior data with length scaling up to thousands. These existing recommendation models, despite effective, are struggling in discovering users' interests in the TIR scenario due to the absence of explicitly modelling the trigger item.

\textbf{Trigger-Induced Recommendation:}
The most relevant work to ours is the R3S \cite{xie2021real}, $i.e.$, Real-time Relevant Recommendation Suggestion, which can be regarded as a research for TIR to some extent. R3S proposes a novel recommendation suggestion task for extended reading in recommendation, aiming to predict users’ intent on extended reading and recommend appropriate relevant items given the current clicked item. Following the definition of our work, the item a user has just clicked can be regarded as the trigger item. Specifically, R3S extract users' interests from multiple aspects including feature interactions, semantic similarity and information gain between clicked item (trigger item) and relevant candidate items. However, R3S fails to capture users' interests from their historical behaviors given clicked item (trigger item) and candidate items. Alternatively, in our DIHN, we devise a FEM module to fuse trigger item and target item embeddings seamlessly, which is further utilized to extract users' interests from their behaviors.

\section{Proposed method}
\label{sec:method}

\begin{figure*}
  \centering
  \includegraphics[width=\textwidth]{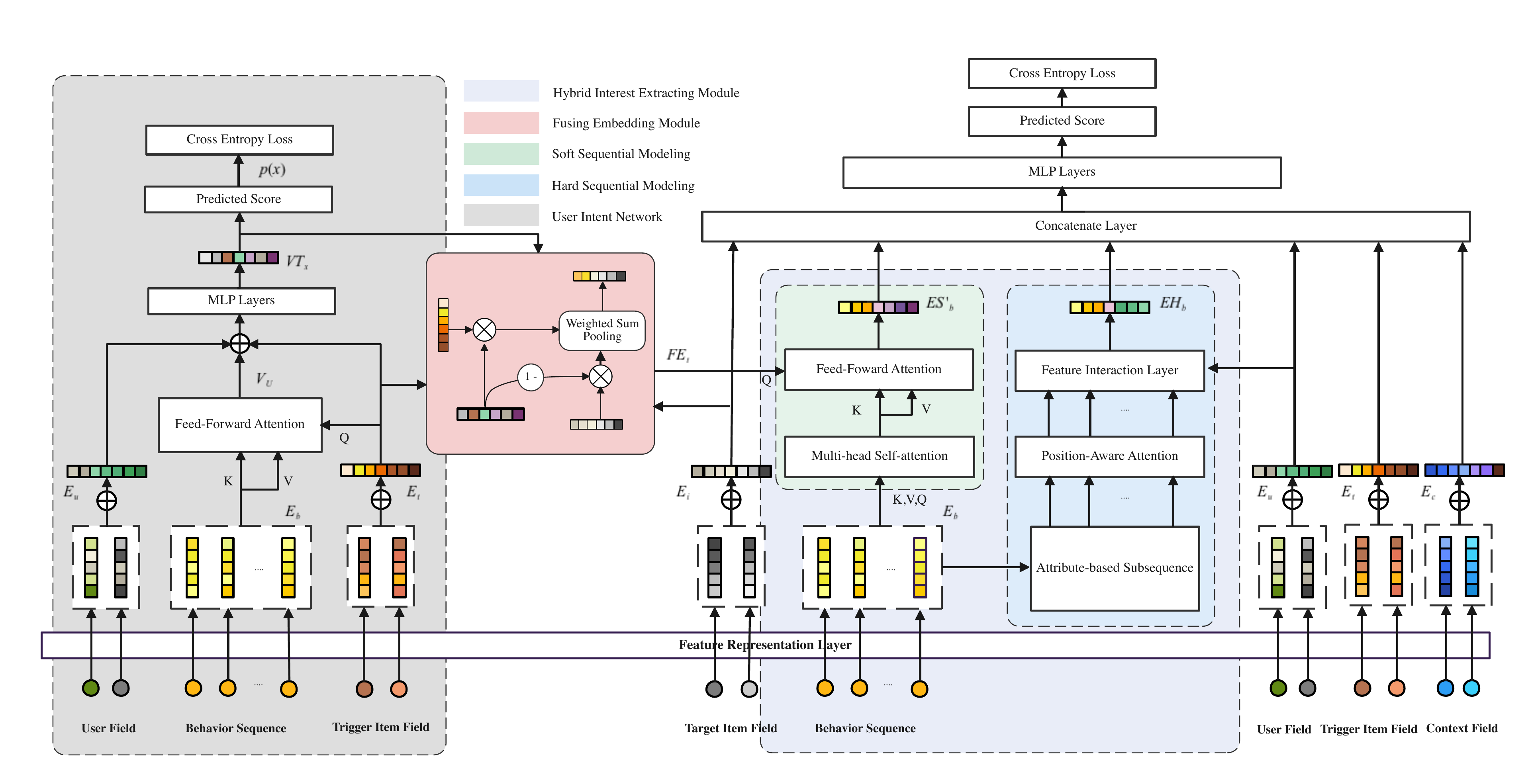}
  \caption{The overview architecture of our proposed DIHN model, which consists of Feature Representation Layer (FRL), User Intent Network (UIN), Fusion Embedding Module (FEM), and Hybrid Interest Extracting Module (HIEM) modules.
  }
  \label{fig:model_pic}
\end{figure*}

In this paper, we propose a novel recommendation model named DIHN for CTR prediction in TIR scenarios. The overall architecture of DINH is depicted in Fig. \ref{fig:model_pic}, which consists of four main components including 1) Feature Representation Layer (FRL), which transforms all kinds of high dimension sparse one-hot vectors into fixed-length low dimension dense vectors; 2) User Intent Network (UIN), which responds to generate a precise probability score to predict user’s intent on the trigger item; 3) Fusion Embedding Module (FEM), which adaptively fuses trigger item and target item embeddings based on the result of UIN; and (4) Hybrid Interest Extracting Module (HIEM), which can effectively highlight users’ instant interest from their behaviors based on the result of FEM. Finally, all the representation features are concatenated with several raw context features and fed into multi-layer perceptron (MLP) layers for final CTR prediction. In the remaining of this section, we will introduce them in detail. 

\subsection{Motivation}
\label{subsec:motivation}

Generally speaking, users could show multiple interests from their historical behaviors, such as electronics, clothing, and snacks, resulting in diverse items recommended accordingly in non-TIR scenarios, $e.g.$, the scenario in Fig. \ref{fig:scene_info}(a). 
Several sequential modelling methods are proposed to capture user's dynamic interests from their historical behaviors with respect to different target items. For example, Deep Interest Network (DIN) \cite{zhou2018deep}, as the excellent representative work, is designed to activate relevant users' behaviors with respect to corresponding targets and obtain adaptive representation vectors for users' interest extraction. Despite effective, we figure out that those existing methods are not so effective in a trigger-induced recommender due to the absence of explicitly modelling the instant interest induced by a trigger item. For example, given the category of the clicked trigger item \emph{Electronic}, it implies that the user is only interested in items related to \emph{Electronic} category at that transient moment. If directly employing DIN without explicitly modeling users' instant interest, the performance of the model will be seriously degraded. Alternatively, we can leverage the trigger item together with target items, to suitably model users' sequential behaviors and explore their exact interests. However, there are several challenges as introduced in previous section should be carefully addressed. In the following, we will introduce how to address these challenges with the elaborate designed components in DIHN.

\subsection{Feature Representation Layer}
\label{subsubsec:feature_representation_layer}

In DIHN, we mainly employ four categories of features namely \emph{User Profile}, \emph{User Behaviors}, \emph{Trigger}, \emph{Target Item}, and \emph{Context} for users' interest representation, where each category feature consists of several fields. For example, \emph{User Profile} contains \emph{age}, \emph{sex}, \emph{purchase level}, $etc$. \emph{User Behaviors} contains the sequential list of users visiting items. \emph{Trigger} as well as \emph{Target Item} contain \emph{Id}, \emph{Category}, $etc$. And \emph{Context} contains \emph{time}, \emph{weather}, $etc$. In addition, feature in each field is normally transformed into high-dimensional sparse one-hot features via encoding. For example, the feature value \emph{male} from \emph{sex} field of \emph{User Profile} category is encoded as [0,1]. Assuming the concatenation results of different fields' one-hot vectors from \emph{User Profile}, \emph{User Behaviors}, \emph{Trigger}, \emph{Target Item} and \emph{Context} are denoted as $X_{u}$, $X_{b}$, $X_{t}$, $X_{i}$, and $X_{c}$, respectively, they can be further transformed into low dimensional dense representations by utilizing embedding layers, named as $E_{u}$, $E_{b}$, $E_{t}$, $E_{i}$, and $E_{c}$, respectively. Besides, in the sequential CTR prediction task, $X_{b}$ usually contains a list of users behaviors, mathematically denoted as $X_{b} =[b_{1};b_{2};...;b_{T}]\in \mathbb{R}^{K\times T} ,b_{t}\in \left \{ 0,1 \right \}^{K}$, where $T$, $K$, and $b_{t}$ represent the length of users' behaviors, the total number of all candidate items, and the one-hot vector of the $t$-th behavior, respectively. Similarly, $E_{b}$ is denoted as $E_{b} =[e_{1};e_{2};...e_{T}]\in \mathbb{R} ^{D\times T}$, where $D$ and $e_{t}$ represent the dimension of dense feature transformed by embedding layer, and the transformed embedding feature of the $t$-th behavior, respectively.

\subsection{User Intent Network}
\label{subsec:intention_network}

In TIR scenarios, users can explicitly express their instant interests via the trigger item. However, users' instant interest induced from a trigger item are inherently noisy, because there are some accidental clicks on wrong items in users behaviors. Therefore, how to evaluate user' real intent on the trigger item remains challenging. Here, we proposed a User Intent Network (UIN) to address the challenge, which responds to generate a precise probability score accounting for user's real intent on the trigger item. 

Referring to the UIN module in Fig.\ref{fig:model_pic}, where we utilize three categories of features, $i.e$, \emph{User Profile}, \emph{User Behaviors}, \emph{Trigger} to estimate the probability. To adaptively calculate user sequential representation of \emph{User Behaviors} with respect to the trigger item, we refer to the architecture of DIN \cite{zhou2018deep}. It can be formulated as:
\begin{equation}
    V_{U} =f(E_{b};E_{t})=\sum_{i=1}^{T}a(e_{i},E_{t})e_{i}= \sum_{i=1}^{T}w_{i}e_{i},
\label{eq:din}
\end{equation}
where $V_{U}$ denotes the user representation feature with respect to $E_{t}$, and $a(.)$ is a feed-forward network whose output is the activation weight $w_{i}$, as illustrated in DIN \cite{zhou2018deep}.

Now, given the user representation feature $V_{U}$, we firstly concatenate it with other dense representation vector $E_{u}$ and $E_{t}$, $i.e.$, $x=[V_{U};E_{u};E_{t}]$. It is then fed into multiple fully connected layers to further learn the high-order feature interactions. Next, after the activation function, we obtain the output of the UIN, denoted as $p(x)$, which represents the predicted probability of the trigger item being clicked in TIR scenarios ($e.g.$, depicted in Fig.\ref{fig:scene_info}(b)). Finally, we define the objective function of this module as follows:
\begin{equation}
    Loss_{t}=-\frac{1}{N} \sum_{(x,y)\in S}(ylogp(x)+(1-y)log(1-p(x))),
\label{eq:din_loss}
\end{equation}
where $S$ denotes the training set with total size $N$, and $y\in \left \{ 0,1 \right \} $ is the ground truth label representing whether users clicking the trigger item or not. Additionally, in TIR scenarios, each sample has several additional information from a trigger item, including features of the trigger item, the label of whether the trigger item being clicked. Referring to the IFFP depicted in Fig. \ref{fig:scene_info}(b), if users clicking the trigger item, the samples from the same IFFP with the trigger item will have the same auxiliary label ($e.g.$, positive), which can be used to supervise the learning of UIN. 

Besides, the representation vector from last MLP layer is denoted as $VT_{x}$, which has the same dimension with $E_{t}$ and $E_{i}$, will be utilized in following module.

\label{subsec:main_method}
\subsection{Fusion Embedding Module}
\label{subsubsec:fusing_embedding}

In non-TIR scenarios, existing representative methods, $e.g.$, DIN \cite{zhou2018deep}, always employ attention mechanism to dynamically re-weigh users historical behaviors with respect to target items. However, in TIR scenarios, we figure out that if directly employing DIN without explicitly modeling users' instant interest induced from the trigger item, the performance of the model will be seriously degraded. Alternatively, a suitable users' behavior modelling solution could be adaptively fuse the trigger item with target item. Specifically, we proposed a Fusion Embedding Module (FEM), which adaptively fuses trigger item and target item embeddings based on the result of UIN. For example with two extreme cases, one is that if users have no any interest on trigger item but only click it casually, the results of FEM module should degrade to the target item embedding, which indicates only behaviors related to the target item will be selected to extract users' interest. The other is that if users have exact interest on the trigger item, the FEM module will degrade to the trigger item embedding, indicating only behaviors related to the trigger item will be selected for users' interest extraction. Actually, the result predicted by UIN reflects the intensity of users' instant interest, which is used ($e.g.$, element-wise products) in FEM to fuse both embeddings in an adaptive manner for better users' interest extraction. It is formulated as:
\begin{equation}
FE_{t}=VT_{x}E_{t}+(\textbf{1}-VT_{x})E_{i},
\label{eq:fem_fusion}
\end{equation}
where $FE_{t}$, $VT_{x}$, $E_{t}$, and $E_{i}$ represent the fusion embedding, the representation vector from last MLP layer in UIN, the embedding of trigger item, and the embedding of target item, respectively. Here, \textbf{1} is a vector with the same dimension as $VT_{x}$, $E_{t}$, and $E_{i}$.


\subsection{Hybrid Interest Extracting Module}
\label{subsubsec:interest_extractor}

In TIR scenarios, users usually show their instant interest on the trigger item. How to extract user's interest from their historical behaviors with respect to the trigger item and the target item simultaneously is unexplored. In this section, given users' instant interest via a trigger item and the fusing result from FEM, we devise a novel module named Hybrid Interest Extracting Module (HIEM), which can effectively highlight users' exact interest from their behaviors. Specifically, from two different behavior modelling perspectives, we proposed the Hard Sequential Modelling (HSM) and Soft Sequential Modelling (SSM), 
which are detailed as follows.

\subsubsection{Hard Sequential Modelling} The trigger item can reflect users' instant interests. For example, given the category of the trigger item \emph{Electronic}, it implies that the user is only interested in items related to \emph{Electronic} category at that transient moment. Following the hard-search mode from SIM \cite{pi2020search}, we propose the Hard Sequential Modelling (HSM), indicating that only behaviors with the same attribute ($e.g.$, category, destination) as the trigger item will be firstly selected and aggregated as a sub behavior sequence, followed by employing it to capture users' corresponding interests.


Taking the attribute \emph{category} as example, mathematically, let $E_{bc}=[e_{1_{c}};e_{2_{c}};...e_{T_{c}}]\in \mathbb{R} ^{D\times T_{c}}$ denotes the sub-sequence from $E_{b} =[e_{1};e_{2};...e_{T}]\in \mathbb{R}^{D\times T}$, each element within $E_{bc}$ has the same category as the trigger item and ordered by their occurring time. As we all known, user's interest may change with time, where more recent behaviors reflect user’s temporal interest better, we apply an attention mechanism with positional encoding as query to adaptively learn the weight for each behavior, where the position of user behavior is the serial number in the behavior sequence. It can be formulated as follows:
\begin{equation}
\alpha _{t} = z^{T}tanh(W_{p}p_{t}+W_{e}e_{t}+b),
\label{eq:hard1}
\end{equation}

\begin{equation}
\alpha_{t} =  \frac{exp(\alpha_{t})}{\sum_{i=1}^{T}exp(\alpha_{i})},  
\label{eq:hard2}
\end{equation}
where $p_{t}\in \mathbb{R} ^{d_{p}}$ is the t-th position embedding, $e_{t}\in \mathbb{R} ^{d_{e}}$ is the feature vector for the t-th behavior, $W_{p}\in \mathbb{R} ^{d_{h} \times d_{p}}$, $W_{e}\in \mathbb{R} ^{d_{h} \times d_{e}}$, $b\in \mathbb{R} ^{d_{h}}$, and $z\in \mathbb{R} ^{d_{h}}$ are learnable parameters, and $\alpha _{t}$ is the normalized weight for the t-th behavior. By weighted-sum pooling, the feature vector list of user's sub behaviors is mapped into fixed-length user representation vector $u_{c}$:
\begin{equation}
u_{c}  =  \sum_{i=1}^{T}\alpha _{i}e_{i}.
\label{eq:hard3}
\end{equation}

In the same way, the above processing method can be applied in other attributes ($i.e.$, destination, tag) of trigger item, which reflect users' interests from different aspects, getting the \emph{destination} based vector $u_{d}$ and \emph{tag} based representation $u_{t}$, respectively. For further capturing the high-order interaction between these representation vectors and original user profile features, we propose the Feature Interaction Module (FIM). To be specific, we first concatenate features into one representation vector as:
\begin{equation}
r_{g} = [E_{u}, u_{c}, u_{d}, u_{t}].
\label{eq:hard3}
\end{equation}

Then, the combination of all the features is taken as the input of the FIM, which utilizes the global attention unit to extract the relationship among different parts of the input features. The output is calculated by:
\begin{equation}
EH_{b} =  \sum_{l=1}^{n}\frac{exp(tanh(r_{l}\cdot W_{l}+b_{l})}{\sum_{l=1}^{L}exp(tanh(r_{l}\cdot W_{l}+b_{l})}r_{l},
\label{eq:hard3}
\end{equation}
where $W_{l}$ and $b_{l}$ are weight and bias matrix, respectively, $r_{l}$ is the representation of each individual feature, $L$ is the number of total features in $r_{g}$, and $EH_{b}$ is the output vector of FIM.

\subsubsection{Soft Sequential Modelling} Different from users' interest extraction in non-TIR scenarios that directly calculating the relevant weight between target item and users' behaviors, while in TIR scenarios, users usually can show their instant interest with the clicked trigger item. An alternative modeling solution is to extract users' interests with respect to the clicked trigger item and the target item simultaneously. Therefore, from another modelling perspective, we proposed the Soft Sequential Modelling (SSM), which adaptively calculates the representation vector of users' behaviors with respect to the fusing result from FEM. In addition, since attention mechanism, specifically the Multi-Head Self-Attention (MHSA), which can capture the dependencies between representation pairs despite their distance within the sequence, has become a key ingredient for sequential modeling, we also adopt the MHSA to effectively extract users' interest representation. Mathematically, it is defined as:
\begin{equation}
   MHSA(F^{l})=[head_{1},head_{2},...,head_{h}]W^{O},
\label{eq:lstm_form5}
\end{equation}

\begin{equation}
    head_{i}=Attention(Q,K,V)=softmax(\frac{QK^{T} }{\sqrt{d/h} }V ),
\label{eq:lstm_form5}
\end{equation}
where $F^{l}$ denotes the $l$-th layer input and $Q=F^{l}W_{i}^{Q},K=F^{l}W_{i}^{K},V=F^{l}W_{i}^{V}$ denotes the linear transformations of the input $F_{l}$. The projection matrix $W_{i}^{Q}\in \mathbb{R} ^{D\times D/h} ,W_{i}^{K}\in \mathbb{R} ^{D\times D/h},W_{i}^{V}\in \mathbb{R} ^{D\times D/h}$ and $W_{O}\in \mathbb{R} ^{D\times D} $ represents the corresponding learnable parameters for each head. $\sqrt{d/h}$ is the scale factor for normalization.

Next, we endow the non-linearity of the self-attention block by applying a feed-forward network, $i.e.$,
\begin{equation}
    F^{l}=[FFN(F_{1}^{l} )^{T};FFN(F_{2}^{l} )^{T};...; FFN(F_{n}^{l} )^{T}],
\label{eq:lstm_form5}
\end{equation}

\begin{equation}
    FFN(x)=(Relu(xW_{1}+b_{1}))W_{2}+ b_{2},
\label{eq:lstm_form5}
\end{equation}
where $W_{1},b_{1},W_{2},b_{2}$ are learnable parameters. 

In this way, given users sequential behaviors $E_{b} =[e_{1};e_{2};...;e_{T}]$, where $T$ and $e_{i}$ represent the length of users' behaviors and the embedding of $i$-th user behavior, respectively, we can obtain the representation vectors $E_{b}^{'}  =[e_{1}^{'} ;e_{2}^{'} ;...;e_{T}^{'}]$ by MHSA. Next, given $E_{b}^{'}$ and the fusing result from FEM, we employ DIN \cite{zhou2018deep} to extract the user representation vector, denoted as $ES_{b}^{'}$.

Then, all the dense representation vectors are concatenated with the raw context features, defined as $x^{'}=[ES_{b}^{'};EH_{b};E_{U};E_{t};E_{i}]$, which are then fed into fully connected layers to learn the higher-order feature interactions, followed by a relu function to obtain the predicted probability of the target item being clicked. Similarly, the objective function of this module is the negative log-likelihood function like Eq. \eqref{eq:din_loss}. We denote it as $Loss_{i}$. Finally, the total loss of our proposed DIHN model is formulated as:
\begin{equation}
    Loss=\alpha Loss_{t} + \beta Loss_{i},
\label{eq:lstm_form5}
\end{equation}
where $\alpha$ and $\beta$ are hyper parameters to balance these two losses. 


\section{Experiments}
\label{sec:experiments}

To evaluate the effectiveness of our proposed DIHN, in this section, we conduct extensive experiments to compare it with several representative SOTA methods on both offline datasets and online deployment. We firstly present the details of benchmark dataset and experimental setup, including the offline dataset preparation,  evaluation metrics, and a brief description of representative SOTA methods. Then, the main results and analysis are presented, followed by ablation studies.

\subsection{Dataset and Experimental Setup}
\label{subsec:evaluation_setting}


To the best of our knowledge, there are no TIR scenario based public datasets tailored for the new proposed problem. To fill this gap, we establish an offline dataset by collecting data from our real-world e-commerce TIR scenarios, named Industrial Dataset. Besides, to demonstrate the generalization of our proposed model, we also carry out experiments on a public dataset, $i.e.$, Alimama Dataset, where we manually construct the trigger items to suit for the TIR problem. Table~\ref{tab:stats} shows the statistics of all datasets. We will introduce them in detail.

\textbf{Industrial Dataset.} 
We firstly establish the offline dataset by collecting the users' behaviors and feedback logs from our online e-commerce platform, which is one of the largest third-party retail platforms in the world. 
Each sample contains \emph{User Profile}/\emph{User Behaviors}/\emph{Trigger}/\emph{Target Item}/\emph{Context} features described in previous section, as well as feedback logs ($i.e.$, whether users click the target item or not). 
Besides, \emph{Trigger} also has extra information, including features of the trigger item and the auxiliary label indicating whether the trigger item has been clicked or not in IFFP. Referring to the IFFP depicted in Fig. \ref{fig:scene_info}(b), if users click the trigger item within IFFP, the auxiliary label is set to 1 ($i.e.$, positive), otherwise 0. Samples collected from the same IFFP with the same trigger item 
have the identical auxiliary label, which can be used to supervise the learning of UIN.


\textbf{Public Dataset.} 
Alimama Dataset\footnote{https://tianchi.aliyun.com/dataset/dataDetail?dataId=56} is a public dataset released by Alimama, which is an online advertising platform of China. 
It contains 26 million records from ad display/click logs with 1 million users and 800 thousand ads in 8 days. 
However, it is not collected from TIR scenarios, due to the lack of trigger items. 
Therefore, we manually define the trigger items as follows. 
For each sample (one display of a user and an target ad at time $t$), we search for the latest clicked advertising of the user within 4 hours before $t$.
It is worth noting that not every sample can be associated with a trigger, since users behaviors are sparse. For instance, some users may have no click behavior in the dataset.
So we only select those samples that can be associated with a trigger item. 
If the trigger item has the same category with the target advertising, the auxiliary label is set to 1, otherwise 0.





\begin{table}[htbp]
\small
  \caption{Statistics of the offline datasets.}
  \label{tab:data}
  \begin{tabular}{llll}
    \hline
   Dataset & \#User & \#Item & \#Impression  \\
   \hline
    \midrule
   Industrial. & 22,269,532 & 201,004 & 436,240,250 \\
   Public. & 26,557,962 & 846,812 & 1,366,056 \\
    \hline
  \bottomrule
\end{tabular}
\label{tab:stats}
\end{table}

\textbf{Experimental Setup.} We implement all the competitive methods in TensorFlow using the Adam optimizer with a exponential decay learning rate schedule. The initial learning rate is set as 0.001 and decay rate is set as 0.9. The hyper-parameters $\alpha$ and $\beta$ in Eq.~\eqref{eq:lstm_form5} are set 1 and 0.8, respectively. We take AUC \cite{zhou2018deep} as the main metric to evaluate model's performance, which is widely adopted in the field of CTR prediction task. We run each method 10 times and report the average result.

\subsection{Competitors}
\label{subsubsec:competitive_methods}

To demonstrate the effectiveness and superiority of DIHN, we compare it with several representative SOTA methods in TIR scenarios, which can be grouped into two categories.

Group 1: Methods without explicitly modelling users' instant interest. Classical models are WDL, DeepFM, and DIN, etc., which do not include information from the trigger item. We show them in detail.


\begin{itemize}

    \item \textbf{WDL} \cite{cheng2016wide}: It jointly trains a wide linear model and a deep neural model, which combines the benefits of memorization and generalization for CTR prediction.


    \item \textbf{DeepFM} \cite{guo2017deepfm}: It emphasizes both low- and high-order feature interactions by combining the power of traditional FM module and deep MLP module.

    
    
    \item \textbf{DIN} \cite{zhou2018deep}: It utilizes attention mechanism to activate relevant users’ behaviors with respect to corresponding targets and learns an adaptive representation vector for users’ interests.
    
     \item \textbf{DIEN} \cite{zhou2019deep}: It adopts an interest extractor layer to capture temporal interests from users' historical behaviors and integrates GRUs with attention mechanism for further capturing the involved interests with respect to the target item.
     
     \item \textbf{MIAN} \cite{zhang2021multi}: It contains a multi-interactive layer to capture multiple representations of user preference from sequential behaviors. Besides, it utilizes a global interaction module to learn the high-order interactions and balances the different impacts of multiple features.
     
     
     
\end{itemize}

Group 2: Methods with explicitly modelling users' instant interest. As aforementioned, there are no prior works in TIR, thus we equip some SOTA CTR models with the capacity of utilizing the trigger item for fair and solid comparison.

\begin{itemize}

    
    \item \textbf{DIN+2TA}: It utilizes the attention mechanism to extract users’ interest representation not only with respect to the target item, but also to the trigger item inducing users' explicit instant interests. 
    
     
    \item \textbf{DIEN+2TA}: It captures users' envolved interests from users' historical behaviors 
    by using the similar attention structure in \textbf{DIN+2TA}.

    
     \item \textbf{R3S}\cite{xie2021real}: It extracts users' interest from multiple aspects including feature interactions, semantic similarity and information gain between clicked item and target items.
    
    \item \textbf{R3S+2TA}: Since the model R3S does not utilize users' sequential behaviors, in this model, we add it as input and capture users' interests with respect to the trigger item and the target item simultaneously.

\end{itemize}

\subsection{Main Offline Comparison Results}
\label{subsubsec:main_comparison_results}

\begin{table}[]
\small
\caption{Comparison results of methods in Group 1.}
\begin{center}
\begin{tabular}{lcc}
\hline
Model      & Industrial (mean ± std)    & Public (mean ± std) \\ \hline
WDL  & 0.7231 ± 0.00019 & 0.6351 ± 0.00034  \\
DeepFM     & 0.7271 ± 0.00031 & 0.6359 ± 0.00039 \\
DIN        & 0.7351 ± 0.00021 & 0.6382 ± 0.00024 \\
DIEN       & 0.7364 ± 0.00041 & 0.6388 ± 0.00033\\
MIAN       & 0.7379 ± 0.00038 & 0.6391 ± 0.00039 \\
\textbf{DIHN}       & \textbf{0.7506 ± 0.00025} & \textbf{0.6421 ± 0.00019}  \\ \hline
\end{tabular}
\end{center}
\end{table}

\begin{table}[]
\small
\caption{Comparison results of methods in Group 2.}
\begin{center}
\begin{tabular}{lcc}
\hline
Model    & Industrial (mean ± std)    & Public (mean ± std) \\ \hline
DIN+2TA  & 0.7401 ± 0.00029 & 0.6394 ± 0.00029 \\
DIEN+2TA & 0.7389 ± 0.00042 & 0.6391 ± 0.00036 \\
R3S      & 0.7378 ± 0.00031 & 0.6392 ± 0.00017 \\
R3S+2TA  & 0.7415 ± 0.00027 & 0.6403 ± 0.00024 \\
\textbf{DIHN}     & \textbf{0.7506 ± 0.00025} & \textbf{0.6421 ± 0.00019}  \\ \hline
\end{tabular}
\end{center}
\end{table}

\subsubsection{The Comparison Results with Competitors in Group 1}
    We start off reporting the AUC of all the competitive methods in Group 1. 
    It can be seen that DeepFM achieves 0.55\% and 0.13\% AUC improvement over the WDL for Industrial and Public datasets, respectively. 
    It demonstrates the remarkable effectiveness of both low- and high-order feature interactions, attributing to the combination of traditional FM and deep MLP modules.
    Different from WDL and DeepFM without employing users' behaviors, DIN, DIEN, and MIAN performance better on both datasets, benefiting from capturing users' interests by modelling users' historical behaviors. 
    For example, DIN achieves AUC gains of 1.10\% and 1.66\% over DeepFM and WDL on the Industrial dataset, respectively. 
    As for the MIAN, it can efficiently learn the multiple fine-grained interactions from historical behaviors, user-specific and contextual interactions. 
    While DIEN only focuses on the evolving interests within users' behaviors.
    Consequently, MIAN achieves 0.21\% and 0.05\% improvement over DIEN on the two datasets, respectively. 
    By contrast, our proposed DIHN significantly outperforms above state-of-the-art competitors, which mainly benefits from explicitly modeling users' instant interest induced from the trigger item. 
    Specifically, compared with MIAN, the improvement on AUC is 1.72\% and 0.47\% on the two datasets, respectively. 
    It is worth mentioning that the gain of 0.01 on the offline dataset always means significant increment for online CTR tasks~\cite{wen2019multi}.
    The experimental comparison results also reveal that explicitly utilizing users' instant interests in TIR scenarios can achieve the best CTR performance. 

\subsubsection{The Comparison Results with Competitors in Group 2}    
    For the competitive methods in Group 2, we utilizes the attention mechanism to activate relevant users’ behaviors not only with respect to target item but also the trigger item. First, DIN+2TA (or DIEN+2TA) simply employs two attention mechanism with respect to the trigger item and target items individually, while R3S+2TA employs a fusion of them. Specifically, R3S+2TA considers semantic relevance and information gain between the trigger item and target items, resulting in 0.19\% (or 0.35\%) and 0.14\% (or 0.18\%) AUC gains over DIN+2TA (or DIEN+2TA) on the two datasets, respectively. Despite effective, it has no elaborated modelling between users' historical behaviors and the fusion results. As for the proposed DIHN, it uses the FEM module to supervise the extraction of users' interests and the HIEM module to highlight users’exact interest from their behaviors, achieving the best performance among all the competitors. For example, the AUC gains over R3S+2TA are 1.23\% and 0.28\% on the two datasets, respectively. It is worth noting that the improvement on the Alimama dataset is not as obvious as on the Industry dataset due to the reason that the Alimama dataset is not directly collected from online TIR scenario but established in a synthetic way where we manually mine the information of trigger items according to self-defined rules. Consequently, it may mismatch with the real online TIR scenario and limit our model's performance. Additionally, compared with competitors in Group 1, which are not explicitly modelling users' instant interest, DIHN also achieves significant gains, which further demonstrates the effectiveness of our method.
    
\subsection{Ablation Study}
\label{subsec:ablation_study}
\begin{table}[]
\small
\caption{Results of ablation study on offline datasets.}
\begin{tabular}{lcc}
\hline
Model            & Industrial (mean ± std)    & Public (mean ± std) \\ \hline
DIHN w/o HSM         & 0.7447 ± 0.00009 & 0.6390 ± 0.00020  \\
DIHN(scalar)             & 0.7484 ± 0.00011 & 0.6409 ± 0.00013 \\
DIHN w/o SSM+m       & 0.7443 ± 0.00031  & 0.6388 ± 0.00034 \\
DIHN w/o SSM+target  & 0.7469 ± 0.00013 & 0.6401 ± 0.00022  \\
DIHN w/o SSM+trigger & 0.7475 ± 0.00019 & 0.6403 ± 0.00012 \\
DIHN w/o SSM+concat  & 0.7459 ± 0.00041 & 0.6399 ± 0.00027 \\ 
\textbf{DIHN}                 & \textbf{0.7506 ± 0.00025} & \textbf{0.6421 ± 0.00019}   \\    \hline
\end{tabular}
\end{table}

To investigate the effectiveness of each component in our model, in this subsection, we present ablation studies on the offline datasets. 

\subsubsection{Effectiveness of HSM module.} To evaluate the impact of HSM module for users' interest extraction, we compared DIHN with \emph{DIHN w/o HSM}, indicating DIHN without the HSM module. The results are shown in Table \ref{subsec:ablation_study}. Since HSM module can help filter irrelevant noise so that the model focuses on the most relevant behaviors for users' interest extraction, especially when the users have intense interest on trigger item, it is a straightforward but effective solution to improve the performance.
    
\subsubsection{Effectiveness of UIN module.} The module UIN is designed to predict users' instant interests on the trigger item. We now investigate whether this module can strengthen users' interests representation. To be specific, we evaluate DIHN without the module while employing four kinds of SSM methods, $e.g.$, \emph{DIHN w/o SSM+m}, \emph{DIHN w/o SSM+target}, \emph{DIHN w/o SSM+trigger}, and \emph{DIHN w/o SSM+concat}, denoting that DIN where mean pooling for the representation of users' sequential behaviors, DIN where the query is target item, DIN where the query is the trigger item, DIN where the query is the concatenation of the trigger item and target items, respectively. The offline comparison results in Table \ref{subsec:ablation_study} show the effectiveness of UIN module and the benefit of using supervised auxiliary label from whether users clicking the trigger item. Users' interest on the trigger item obtained by the UIN module can help extract users' interests with respect to the target item and bring further performance improvement.

\subsubsection{Effectiveness of FEM module.} Table \ref{subsec:ablation_study} also shows the results of different embedding fusing methods in the FEM module, where DIHN (scalar) denotes the model with scalar fusing operator, $i.e.$, replacing the vector $VT_{x}$ with a scalar $p(x)$ in Eq.\ref{eq:fem_fusion}, while DIHN uses an element-wise embedding fusing operator. As can be seen, DIHN obtains 0.3\% and 0.19\% gains over DIHN (scalar) on the two offline datasets, respectively. Although DIHN (scalar) can directly employ the predicted score from UIN to control the fusion of two embeddings, it only uses the scalar $p(x)$, which ignores the importance at different dimensions. Therefore, we propose to use the element-wise fusion operator in the FEM module. 

\subsection{Online A/B Testing Results}
\label{subsec:online_ab_testing_results}

To further validate the effectiveness of our proposed model, we also conducted online A/B testing on our e-commerce platform. However, it is not an easy job to deploy the proposed model in our online recommender system since it serves at the scale of tens of millions of users every day. Besides, the traffic is very expensive from the business view. Considering this fact, we only deployed the best offline method as our baseline method, $i.e.$, R3S+2TA model. Simultaneously, to make the online evaluation fair, confident, and comparable, each deployed method for an A/B test has involved the same number of users, $i.e.$, millions of users. Careful online A/B testing was conducted from 2021-06 to 2021-07. DIHN contributes up to 6.5\% CTR promotion, which is a significant improvement and demonstrates the effectiveness of our proposed method. Now, DIHN has been deployed online and serves the main traffic.


\subsection{Case Study}

\begin{figure}
  \centering
  \includegraphics[width=0.95\linewidth]{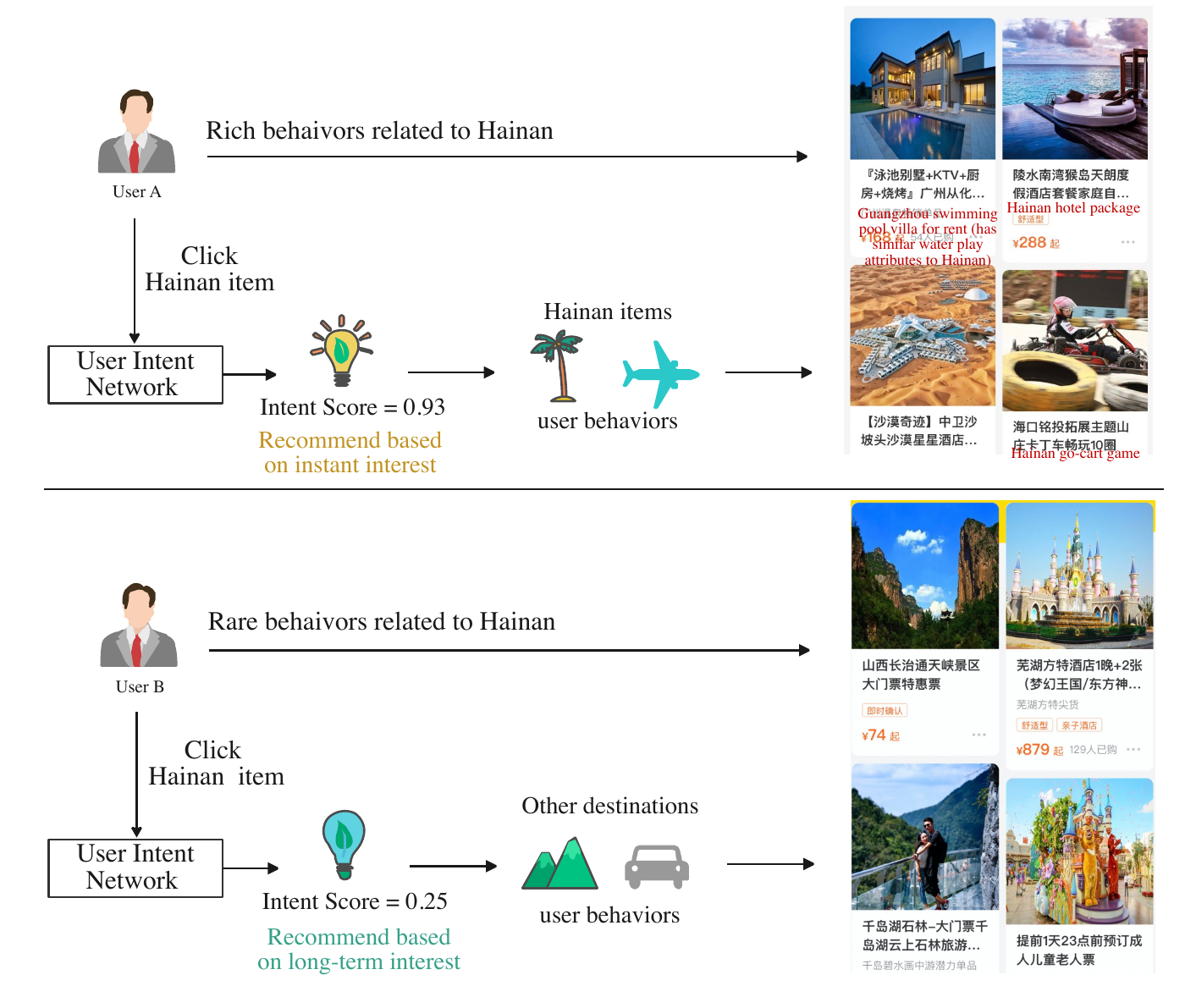}
  \caption{ 
  Two cases of DIHN in the travel scenario, where two different users clicked the same trigger item (Hainan vacation item). User A (top) has rich behaviors related to Hainan, while user B (bottom) has rare behaviors. 
  }
  \label{fig:case_study}
\end{figure}

Taking the travel scene as an example, we present two real-world cases shown in Figure \ref{fig:case_study} to illustrate the effectiveness of DIHN. When user A comes into the TIR scenario, the UIN module predicts A's intent score for the trigger item, $i.e.$, an item from Hainan province, as 0.93, which indicates the user has intense interest on items related to Hainan province. Therefore, more products related to Hainan or similar scenic spots will be recommended to satisfy the user's instant intention, as shown in Figure \ref{fig:case_study}. Different from user A, user B has rare Hainan province related behaviors, therefore the intent score on the trigger item predicted by UIN module is very low, $i.e.$, 0.25. Such condition indicates user B has divergent interests, and the recommender system in IFFP should recommend less items related to Hainan province. As can be seen from the above two cases, DIHN can seamlessly adjust the suitable recommendation mode based on the user's instant interest on the trigger item.

\section{Conclusion}

In this paper, we introduce a new Trigger-Induced Recommendation (TIR) problem, where users' instant interest can be explicitly induced with a trigger item. Due to the discrepancy between TIR scenarios and non-TIR scenarios, we figure out that existing recommendation models are struggling in discovering users’ instant interests in TIR scenarios. To tackle the problem, we propose a novel recommendation method DIHN, which shows great benefits of modeling users’ instant interests for CTR prediction. Using a User Intent Network, DIHN can predict the extent of user's intent on the trigger item and accordingly rely on the trigger item or target item embeddings for user's interest extraction. DIHN not only achieves improvement over representative state-of-the-art methods on offline datasets, but also gains 6.5\% CTR promotion on online e-commerce platform, confirming its superiority for TIR. 

\label{subsec:online_ab_testing_results}

\bibliographystyle{ACM-Reference-Format}
\bibliography{sample-base}

\appendix









\end{document}